# THE POWER SPECTRUM OF GALAXIES IN THE NEARBY UNIVERSE[1]

L. Nicolaci da Costa,[2,3] Michael S. Vogeley,[4] Margaret J. Geller,[5]
John P. Huchra,[5] Changbom Park[6]

[1] Based on observations carried out at the Cerro-Tololo Inter-American Observatory, Complejo Astronómico El Leoncito, European Southern Observatory, F.L. Whipple Observatory, and South African Astronomical Observatory.

[2] Institut d'Astrophysique, 98 bis Boulevard Arago, Paris F75014 France.
Email: ldacosta@iap.fr

[3] Departamento de Astronomia, CNPq/Observatório Nacional,
rua General José Cristino 77, Rio de Janeiro, R.J. 20921 Brazil.

[4] Department of Physics and Astronomy, Johns Hopkins University,
Baltimore, MD 21218.
Email: vogeley@pha.jhu.edu

[5] Harvard-Smithsonian Center for Astrophysics, 60 Garden Street,
Cambridge, MA 02138.
Email: mjg@cfa.harvard.edu, huchra@cfa.harvard.edu

[6] Department of Astronomy, Seoul National University, Seoul, 151 Korea.
Email: cbp@astrograte.snu.ac.kr

astro-ph/9410066  20 Oct 1994

# ABSTRACT

We compute the power spectrum of galaxy density fluctuations in a recently completed redshift survey of optically-selected galaxies in the southern hemisphere (SSRS2). The amplitude and shape of the SSRS2 power spectrum are consistent with results of the Center for Astrophysics redshift survey of the northern hemisphere (CfA2), including the abrupt change of slope on a scale of $30 - 50 h^{-1}$Mpc; these results are reproducible for independent volumes of space and variations are consistent with the errors estimated from mock surveys. Taken together, the SSRS2 and CfA2 form a complete sample of 14,383 galaxies which covers one-third of the sky. The power spectrum of this larger sample continues to rise on scales up to $\sim 200 h^{-1}$Mpc, with weak evidence for flattening on the largest scales. The SSRS2+CfA2 power spectrum and the power spectrum constraints implied by COBE are well-matched by an $\Omega h \sim 0.2$, $\Omega + \lambda_0 = 1$ CDM model with minimal biasing of optically-selected galaxies.

*Subject headings:* cosmology: observations – galaxies: clustering – large-scale structure of universe

## 1. INTRODUCTION

The power spectrum of galaxy density fluctuations (PS hereafter) provides an important constraint on theories for the formation of large-scale structure. The redshift space PS results for a variety of galaxy redshift samples (Baumgart & Fry 1991, Kaiser 1991, Peacock & Nicholson 1991, Park, Gott, & da Costa 1992, Vogeley et al. 1992, Fisher et al. 1993, Feldman, Kaiser & Peacock 1993, Park et al. 1994) roughly follow a power law $P(k) \propto k^n$ with a slope ranging from $n \approx -2$ on small scales ($\lambda \lesssim 30 h^{-1}$ Mpc) to $n \approx -1.1$ on intermediate scales ($30 h^{-1}$ Mpc $< \lambda < 120 h^{-1}$ Mpc). However, the PS shape on larger scales and the overall normalization of the PS differ among these samples. Some authors claim a turnover in the PS (Feldman et al. 1994); others claim a continued rise (Park et al. 1994, PVGH hereafter) up to $200 h^{-1}$Mpc. In addition, different groups find features in the PS at different scales. Discrepancies among PS estimates may arise from differences in sample selection and/or the method of PS analysis, as well as from the uncertainty due to the finite volume of each survey.

Redshift surveys to moderate depth with wide angular coverage and dense sampling like the magnitude-limited CfA2 (Geller & Huchra 1989, Huchra, Vogeley, & Geller 1994) and SSRS2 surveys (da Costa et al. 1994) are suitable for estimating the PS of the galaxy distribution over a large range of scales. Here we use the three independent datasets, SSRS2, CfA2 North, and CfA2 South, to test the reproducibility of the PS, to assess the significance of features, and to estimate the sampling fluctuations directly from the data. We also use the combined SSRS2+CfA2 sample to obtain a new estimate of the PS. The location of the SSRS2, opposite to the CfA2 North sample and contiguous with CfA2 South, provides a longer baseline in three dimensions to probe large wavelength modes.

## 2. SSRS2 AND CfA2 GALAXY SAMPLES

SSRS2 includes 3600 galaxies and is complete over 1.13 steradians of the southern galactic cap in the declination range $-40° \leq \delta \leq -2.5°$ and $b \leq -40°$ (da Costa et al. 1994). CfA2 North covers the declination range $8.5° \leq \delta \leq 44.5°$ and right ascension range $8^h \leq \alpha \leq 17^h$ and includes 6500 galaxies (Geller & Huchra 1989). CfA2 South covers the region $-2.5° \leq \delta \leq 48°$ and $20^h \leq \alpha \leq 4^h$ and includes 4283 galaxies (Huchra et al. 1994). Both SSRS2 and CfA2 are magnitude-limited to $m_{B(0)} \leq 15.5$. SSRS2 is derived from plate scans (see da Costa et al. 1994 for details). CfA2 is based on the Zwicky catalog. The total SSRS2+CfA2 sample includes 14,383 galaxies and covers 4.08 steradians.

We correct the heliocentric velocities $cz$ for Solar motion with respect to the centroid of the Local Group ($\Delta v = 300 \sin(l) \cos(b)$ km s$^{-1}$), transform the radial velocities into comoving coordinate distances, and compute the absolute magnitude $M$ of galaxies in the sample according to the standard equations (1) and (2) in PVGH.

The dense sampling of these redshift surveys allows us to examine volume-limited samples and thereby avoid the complications of correcting for the selection function and the possible dependence on luminosity (and therefore on distance in a flux-limited sample) of the clustering amplitude. We examine volume-limited samples with depth in comoving coordinates of $101h^{-1}$ Mpc and $130h^{-1}$ Mpc, corresponding to absolute magnitude limits $M_B = -19.7 + 5\log h$ and $-20.3 + 5\log h$, respectively. After trimming the boundaries of CfA2 to limit extinction (estimated from Burstein & Heiles 1982) to $\Delta m \lesssim 0.1$ in CfA2 North and $\Delta m < 0.3$ in CfA2 South (see Figure 1 of PVGH), the 101-depth SSRS2, CfA2 North, and CfA2 South sub-samples contain 817, 1031, and 478 galaxies, respectively. The corresponding 130-depth sub-samples contain 422, 385, and 223 galaxies.

## 3. POWER SPECTRUM ESTIMATION

We use the technique described by PVGH to measure the PS directly from the galaxy distribution. Our estimate of the PS is

$$P(\mathbf{k}) = \left(\langle|\delta_{\mathbf{k}}|^2\rangle - \frac{1}{N}\right) \left(\sum_{\mathbf{k}} |W_{\mathbf{k}}|^2\right)^{-1} \left(1 - |W_{\mathbf{k}}|^2\right)^{-1} \quad (1)$$

where

$$\delta_{\mathbf{k}} = \frac{1}{N} \sum_j e^{i\mathbf{k}\cdot\mathbf{x_j}} - W_{\mathbf{k}}, \quad (2)$$

is the Fourier transform of the galaxy distribution minus the Fourier transform of the survey window (where we define the window function $W(\mathbf{x}) = 1$ inside the survey and 0 outside). This method accounts for the contributions from shot noise (1/N term) and the survey geometry, and corrects the normalization (second factor) and shape (third factor) for the finite volume of the survey.

The finite volume of the samples causes a power loss at large wavelengths because equation (2) implicitly assumes that the density of a sample equals the mean density of the universe. The third factor on the right side of equation (1) is an approximate correction for this large-scale power damping (Peacock & Nicholson 1991). To correct more accurately for this effect, we compute the PS for mock redshift surveys drawn from an N-body simulation (CDM with $\Omega h = 0.2$, $b = 1$, $\sigma_8 = 1$; see PVGH for details of this particle-mesh simulation) and compute the ratio of the mock survey PS to the true PS. From this comparison we derive the correction to apply to the estimated PS. To construct each mock survey, we 'observe' the simulation from a different randomly selected 'galaxy' through the same window and with the same sampling density as the data. The corrected PS estimates are reliable up to scales where $|W_k|^2 \lesssim 0.2$. Thus we restrict calculation of the PS to scales $\lambda \lesssim 100 h^{-1}$Mpc for the individual samples (SSRS2, CfA2 North, CfA2 South) and $\lambda \lesssim 200 h^{-1}$Mpc for the combined sample. All results below include this correction.

We estimate the errors in the PS caused by the finite survey volume and sampling density from the variance in the PS of 100 mock surveys drawn from the $\Omega h = 0.2$ CDM simulation. These Monte Carlo uncertainties include the effects of non-linear mode coupling and therefore are larger than those obtained by assuming purely Gaussian fluctuations (cf. Feldman et al. 1994).

## 4. RESULTS

We demonstrate the reproducibility of our PS results by direct comparison of the PS obtained for the SSRS2, CfA2 North, and CfA2 South samples which probe different structures in comparable volumes. Figures 1a and 1b show the PS for each of these samples volume-limited to depths of $101 h^{-1}$Mpc and $130 h^{-1}$Mpc, respectively. The shapes of the PS are remarkably similar for $k > 0.2$. The covariance of the sample-to-sample PS uncertainty is significant (see Figure 2 below); thus the small systematic variation between these three samples, especially at small $k$, is within the $1\sigma$ uncertainties. Furthermore, the shapes of the PS for the shallower CfA1 and SSRS1 surveys (Park, Gott & da Costa 1992) are also consistent with these results. The new, deeper samples probe a volume 4 times larger and thus are nearly independent of the shallower samples. The reproducibility of the PS for independent samples indicates that the PS for optically-selected galaxies is robust for scales $\lambda \lesssim 100 - 150 h^{-1}$Mpc. From the 68% error bars and the scatter among the three independent samples, we conclude that the error estimates derived in section 3 are consistent with the observed sampling fluctuations.

The PS of the 130-depth samples of both CfA2 North and SSRS2 exhibit an abrupt change of slope near $k \sim 0.2$. The reproducibility of this slope change in independent samples strengthens our suggestion that it is a real feature of the galaxy distribution, possibly reflecting the presence of voids $30 - 50 h^{-1}$ Mpc in diameter (e.g., Geller & Huchra 1989, da Costa et al. 1994, Shectman et al. 1992).

In each sample, the overall amplitude of the PS increases by a factor of $\sim 1.4$ from the 101h$^{-1}$ Mpc to the 130h$^{-1}$ Mpc samples. PVGH attribute the amplitude variation in the CfA2 PS to luminosity segregation in the optically-selected CfA2 sample; the larger PS amplitude corresponds to intrinsically brighter galaxies. This possible effect should be taken into account in interpreting the results from flux-limited surveys.

The combined SSRS2+CfA2 sample yields our best constraints to date on the redshift space PS of optically-selected galaxies. Figure 1c shows the PS of CfA2 (combined CfA2 North and South; PVGH) and the combined SSRS2+CfA2 sample. The PS slope is $n \approx$ $-2.1$ on scales up to $\lambda \sim 30h^{-1}$Mpc, then bends to $n \sim -1$ and continues to rise on scales up to $\lambda \sim 200h^{-1}$Mpc. At the largest scale where we compute the PS for the deeper sample, $\lambda = 388h^{-1}$Mpc, the PS appears to turn over. Consistent with the 50% increase of the volume relative to CfA2, there is a significant decrease in the uncertainty of the PS for the smallest wavenumbers (cf. PVGH). The errors are, however, large enough to admit the turnover necessary to match the COBE results at larger scale. We conclude that the SSRS2+CfA2 PS, which is corrected for large-scale damping of the PS due to the survey window, shows no clear evidence for turnover.

The combined SSRS2+CfA2 PS tracks the CfA2 PS remarkably well, including the break in the PS of the deeper samples at $k \sim 0.2$ (this feature may be absent in the shallower samples which contain a smaller number of resolution elements on this scale). We evaluate the significance of this feature by examining the PS of mock SSRS2+CfA2 surveys from the CDM simulations, which contain no intrinsic feature on this scale. In the mock surveys such apparent features occur only a few percent of the time in samples of this size. Figure 2 shows the PS of 6 randomly selected mock SSRS2+CfA2-130 surveys; some of these PS show features similar in amplitude to that seen in the data (compare with Figure 1c). Detection of the $k \sim 0.2$ feature in a still larger volume is required to confirm a departure from the model. This test also demonstrates the large covariance between PS measurements at different $k$; the covariance for the smaller SSRS2, CfA2 North and South samples is even more significant.

Unlike our optical galaxy samples, the PS for different redshift surveys of IRAS galaxies appear to disagree. The IRAS 1.2Jy sample results (Fisher et al. 1993) are similar to ours in shape but lower in amplitude. In contrast, the QDOT PS (Feldman et al. 1994) shows a definite peak at $\lambda \approx 200h^{-1}$Mpc. For samples of the size of SSRS2+CfA2, the mock surveys in Figure 2 indicate that there may be substantial variation at large scale among individual realizations. An estimate of the real-space PS of optically-selected galaxies based on the much larger APM dataset (Baugh & Efstathiou 1993) peaks around $200h^{-1}$Mpc and then flattens on larger scales.

The redshift-space galaxy PS and limits on the mass PS at $z \sim 1000$ inferred from the COBE observations provide strong constraints on cosmological models. Here we bring the COBE data into the *redshift-space* of our PS measurements and compare both CMB and galaxy observations with CDM models. In Figure 1d we plot our best estimate of the PS for optical galaxies (SSRS2+CfA2) along with error boxes for the mass PS that obtain from the COBE DMR experiment for the two CDM models presented (Efstathiou, Bond, & White 1992; Kofman, Gnedin, & Bahcall 1993) with $Q_{rms} = 17.1 \pm 2.9\mu$K (Wright et al. 1994). The amplification of the PS in redshift space is described by Kaiser (1987). We plot redshift-space mass PS (identical to the galaxy PS if unbiased) for CDM with $\Omega = 1$, $h = 0.5$ and CDM with $\Omega = 0.4$, $h = 0.5$, and $\Omega + \lambda_0 = 1$, both normalized to $\sigma_8 = 1$. We compute these PS from particle-mesh N-body simulations (see PVGH for details). With this normalization, these CDM models require no biasing of galaxies vs. mass. Unbiased CDM with $\Omega = 1$ lacks power on scales $\lambda \sim 100h^{-1}$ Mpc. CDM with $\Omega h = 0.2$, $\Omega + \lambda_0 = 1$, and $b \approx 1$ ($\pm 0.2$ for the $1\sigma$ COBE uncertainty) is consistent with both the galaxy PS and COBE. A somewhat better fit would obtain for slightly larger $\Omega h$ (Kofman et al. 1993 fit $\Omega h \approx 0.28$ to the CfA1+2 PS from Vogeley et al. 1992). On scales sampled by the galaxy PS, the PS of an $\Omega h \sim 0.2$ model with $\lambda_0 = 0$ is nearly identical to that of a model with $\lambda_0 = 1 - \Omega$: either type of model is consistent with the SSRS2+CfA2 PS. However, a non-zero cosmological constant strongly affects the the amplitude of mass fluctuations at the present epoch implied by COBE (e.g., Kofman et al. 1993).

## 5. CONCLUSIONS

The independent optically selected samples, SSRS2, CfA2 North, and CfA2 South, demonstrate the reproducibility of the PS in different directions of the sky, and yield a direct estimate of the sampling errors consistent with mock surveys drawn from N-body simulations. The combined CfA2+SSRS2 sample is the only sample currently available with a sufficient number of galaxies to allow the calculation of the PS for volume-limited samples. These samples are large enough to begin to evaluate departures of the PS from power law behavior in the range 30–50 $h^{-1}$ Mpc.

In spite of the detailed differences in sample selection, the PS for different optically selected surveys with $m_{B(0)} = 15.5$ surveys are within $1\sigma$. Both the shape and amplitude on scales up to $\lambda \approx 150 h^{-1}$ Mpc are reproducible for independent regions which sample different structures. The feature at $k \approx 0.2$ h Mpc$^{-1}$, first noted by Vogeley et al. (1992), appears in both the SSRS2 and CfA2 samples, and thus may be a real feature of the galaxy distribution. Baugh & Efstathiou (1993) report a feature at slightly smaller wavenumber in the *real-space* PS derived from projected counts.

Combining the SSRS2 with the similarly-selected CfA2 sample to reduce the uncertainty on large scales, we obtain our best estimate of the redshift-space PS of optically-selected galaxies. The PS continues to rise on scales up to $\lambda \approx 200 h^{-1}$ Mpc. There is a suggestion of flattening of the spectrum on large scales, but the evidence is marginal because the smearing effects of the window function and normalization problems are the most severe at the largest wavelengths. The shape and amplitude of the SSRS2+CfA2 PS agree with the results for the CfA2 survey alone (PVGH).

In agreement with Kofman et al. (1993) and PVGH, the CDM models that best match our galaxy PS estimates and the limits on the mass PS from the COBE have minimal biasing ($b \approx 1$), $\Omega h \sim 0.2$, and $\Omega + \lambda_0 = 1$. Note that our galaxy PS alone is relatively insensitive to $\lambda_0$, thus an open universe CDM model is also admitted by the galaxy data.

The decrease in the PS uncertainty in the combined SSRS2+CfA2 sample suggests that further surveys of independent regions of the local ($z < 0.05$) universe will improve our estimation of the galaxy PS on scales up to $\lambda \approx 200 h^{-1}$ Mpc. Such surveys would also be particularly useful in determining the reality of the feature indicated at $k \sim 0.2$. Confirmation of this feature from a third sample similar to SSRS2 would be a statistically significant departure from current CDM models. To achieve these gains, we are extending the $m_B \leq 15.5$ survey to cover the northern galactic cap for declinations $\delta \leq 0°$.

Acknowledgments

LNdC, MJG and JPH thank their collaborators in the SSRS2 for the use of the data prior to publication. LNdC thanks the John Simon Guggenheim Foundation and the Harvard-Smithsonian Center for Astrophysics for generous support. MSV is supported in part by the Sloan Digital Sky Survey and NSF grant 9020380; MJG and JPH by NASA grant NAGW-201 and the Smithsonian Institution; and CBP by the S.N.U. Daewoo Research Fund.

# FIGURE CAPTIONS

Figure 1 - PS of SSRS2 and CfA2 redshift samples.

*(a)* Comparison of the PS measured for samples of SSRS2, CfA2 North, and CfA2 South, volume limited at $101h^{-1}$ Mpc. Error bars on SSRS2 are 68% confidence limits computed for mock surveys drawn from an $\Omega h = 0.2$ CDM simulation. The variation of the PS among the individual samples is consistent with this uncertainty.

*(b)* Comparison of the PS measured for samples of SSRS2, CfA2 North, and CfA2 South, volume limited at $130h^{-1}$ Mpc. Error bars are as in 1a. The change in slope at $k \sim 0.2$ occurs in all three samples, suggesting that this is a genuine feature of the galaxy distribution. Comparison of (a) and (b) shows that, for each of the three samples, the PS amplitude of the 130 depth sub-sample is $\sim 40\%$ larger than the corresponding 101 depth sub-sample.

*(c)* PS of the combined SSRS2+CfA2 sample and CfA2 only. Open and filled symbols shows the PS for samples volume-limited at $101h^{-1}$ Mpc and $130h^{-1}$ Mpc comoving distance, respectively. The addition of SSRS2 to CfA2 yields a PS nearly identical to the CfA2 PS, but with smaller uncertainty.

*(d)* Comparison of the PS of the combined SSRS2+CfA2 sample (symbols as in 1c) with limits on the mass PS from COBE and PS of two CDM models, all presented in *redshift space*. Upper solid line is the PS of CDM with $\Omega h = 0.2$ and $\Omega + \lambda_0 = 1$. Lower solid line is the PS of CDM with $\Omega h = 0.5$. We compute these PS from particle-mesh N-body simulations of these models. Both models are unbiased and normalized to $\sigma_8 = 1$. Dashed lines show the PS of these models on linear scales. The boxes indicate constraints ($1\sigma$) on the redshift-space mass PS from the COBE DMR for these two models.

Figure 2 - PS of six mock SSRS2+CfA2 surveys, volume limited at $130h^{-1}$ Mpc, drawn from the $\Omega h = 0.2$ CDM simulation (filled squares). Solid line shows the mean value of the redshift-space PS for this model. The large variation of the PS for $k < 0.05$ demonstrates the uncertainty in the detection of a turnover of the PS. Individual realizations such as these also show features similar to the $k \sim 0.2$ feature in the observed PS. This test also demonstrates the strong covariance between PS estimates at different wavenumber caused by convolution of the true PS with the window function of the survey.

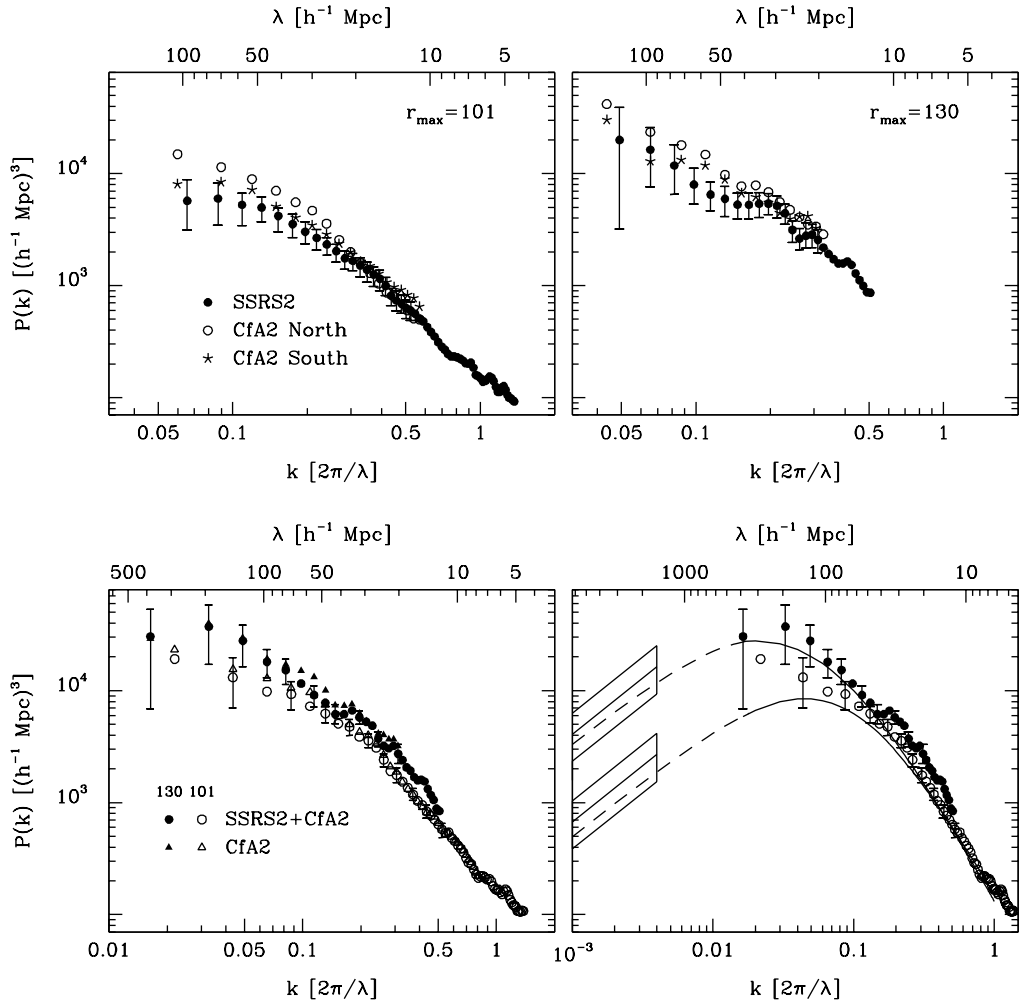

FIGURE 1

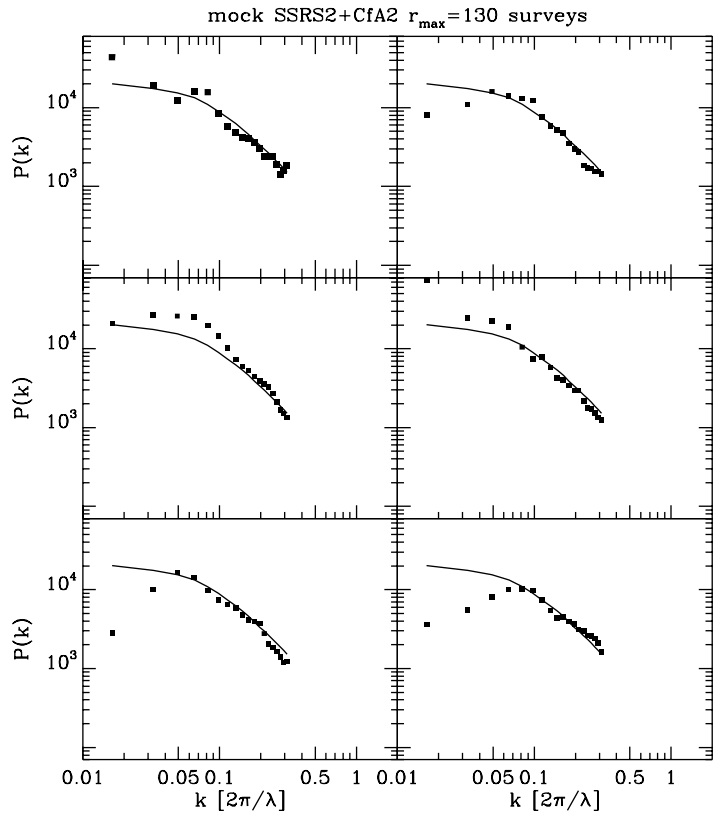

FIGURE 2